\documentclass[12pt]{article}
\usepackage{epsfig}
\begin{document}
\begin{titlepage}
\begin{center}


\hfill{APCTP/05-009}

\vspace{2cm}

{\Large \bf
A QCD sum rule study of the light scalar mesons} \\
\vspace{0.50cm}
\renewcommand{\thefootnote}{\fnsymbol{footnote}}
Hee-Jung Lee \footnote{hjlee@www.apctp.org} \vspace{0.50cm}

{\it Asia Pacific Center for Theoretical Physics, POSTECH, Pohang
790-784, Korea} \vskip 1ex

\end{center}
\vskip 0.5cm

\centerline{\bf Abstract} We examine the interpretation of the
light scalar meson nonet as bound states of the scalar diquark and
the scalar antidiquark using the QCD sum rule approach. Our
results are obtained by means of the operator product expansion
(OPE) including operators up to dimension 8. They show no evidence
of the coupling of the tetraquark states to the light scalar meson
nonet.
\vskip 0.3cm
\leftline{Pacs: 11.55.Hx, 12.38.Aw, 12.38.Lg}
\leftline{Keywords: scalar meson, diquark, QCD sum rule, OPE, tetraquark}

\vspace{1cm}

\end{titlepage}

\setcounter{footnote}{0}
\section{Introduction}

The quasi--bound scalar diquark is one of the main candidates as a
building block of stable multiquark systems \cite{jaffe}. Both,
perturbative one--gluon exchange \cite{jaffe2} and
non--perturbative instanton dynamics \cite{shuryak} favor the
existence of such clusters inside conventional and exotic hadrons.
Maiani et al~\cite{Maiani04} associate the unusual properties of
the light scalar nonet of mesons $\sigma(600)$, $\kappa(800)$,
$f_0(980)$ and $a_0(980)$ to their structure as bound states of
diquark and antidiquark.

It is important to obtain a justification for such tetraquark
picture from QCD. Recently, the QCD sum rule
techniques~\cite{QCDSR} were used in papers by Brito et
al~\cite{Brito05} and Wang et al~\cite{chinese05} to calculate the
decays and masses of the members of the light scalar meson nonet
in a tetraquark picture. Their calculation took only into
account the contributions from operators up to dimension $d=6$ in
the OPE. In~\cite{Lee05} it was shown that, for multiquark
systems, potentially important contributions to the QCD sum rules
may arise from the operators of higher dimension $d>6$ and, if not
considered, wrong conclusions about the properties of the exotic
hadrons might be drawn.

In this Letter we apply the QCD sum rules (SR) technique to the
light scalar meson nonet described as systems composed of the
scalar diquark and the scalar antidiquark as done
in~\cite{Brito05}. Using the factorization hypothesis for
calculating the OPE, we show that the contribution from the
operators of dimension 8 is dominant and leads to the destruction
of SR. We find no evidence for the coupling of the above structure
of diquarks to the light scalar meson nonet within the QCD SR
approach.

\section{QCD sum rules for scalar nonet}

In the picture of the tetraquark states, the light scalar nonet
is generated by the diquark in the $\bar{\bf 3}_f$ and the antidiquark
in the ${\bf 3}_f$, where the subscript $f$ stands for flavor.
The diquark and the antidiquark are assumed to belong to $\bar{\bf 3}_c$,
${\bf 3}_c$ in color space and to spin zero state.
Their conventional wave functions in flavor space are given by
\begin{eqnarray}
&&\sigma(600)=[ud][\bar{u}\bar{d}],\ \ \
f_0(980)=\frac{1}{\sqrt{2}}\bigg([us][\bar{u}\bar{s}]
+[ds][\bar{d}\bar{s}]\bigg),\ \ \
\nonumber\\
&&a_0^+(980)=[us][\bar{d}\bar{s}], \ \ \
a_0^0=\frac{1}{\sqrt{2}}\bigg([us][\bar{u}\bar{s}]-[ds][\bar{d}\bar{s}]\bigg), \ \ \
a_0^-=[ds][\bar{u}\bar{s}],
\nonumber\\
&&\kappa^+(800)=[ud][\bar{d}\bar{s}], \ \ \ \kappa^0=[ud][\bar{u}\bar{s}], \ \ \
\bar{\kappa}^0=[us][\bar{u}\bar{d}],\ \ \ \kappa^-=[ds][\bar{u}\bar{d}],
\label{wff}
\end{eqnarray}
where the square bracket represents the normalized antisymmetric
diquark (antidiquark) state~\cite{jaffe}.

From this structure, the interpolating current for the scalar
nonet in Eq.~(\ref{wff}) can be written as
\begin{equation}
J_S=N_S\epsilon_{abc}\epsilon_{ade}(q_{1b}^{T}\Gamma q_{2c})
(\bar{q}_{3d}\tilde{\Gamma} \bar{q}_{4e}^T)\ ,
\end{equation}
where $\tilde{\Gamma}=\gamma^0\Gamma^\dagger\gamma^0$ and $N_S$ is
the normalization constant. Here the indices $a, b, c,\cdots$
denote color and the subscripts $1,2,3,4$ are introduced for
flavor. The index $S$ labels each meson in the scalar nonet.
$\epsilon_{abc}$ and $\epsilon_{ade}$ guarantee that the diquark
and the antidiquark belong to $\bar{\bf 3}_c$ and ${\bf 3}_c$,
respectively. The antisymmetric structure of the nonet in both
flavor and color space requires that the spin matrix $\Gamma$ must
have the following property
\begin{equation}
\Gamma^T=-\Gamma
\end{equation}
under the transpose of the spin indices. Here we take $\Gamma=C\gamma_5$
in order to consider the scalar diquark--antidiquark system.
The interpolating currents for each meson in the nonet read
\begin{eqnarray}
J_{\sigma}&=&\epsilon_{abc}\epsilon_{ade}
(u_b^TC\gamma_5d_c)(\bar{u}_dC\gamma_5 \bar{d}_e)\ ,
\nonumber\\
J_{f_0}&=&\frac{1}{\sqrt{2}}\epsilon_{abc}\epsilon_{ade}\bigg(
(u_b^TC\gamma_5 s_c)(\bar{u}_dC\gamma_5 \bar{s}_e)+(u\rightarrow d)\bigg)\ ,
\nonumber\\
J_{a_0^0}&=&\frac{1}{\sqrt{2}}\epsilon_{abc}\epsilon_{ade}\bigg(
(u_b^TC\gamma_5 s_c)(\bar{u}_dC\gamma_5 \bar{s}_e)-(u\rightarrow d)\bigg)\ ,
\nonumber\\
J_{\kappa^+}&=&\epsilon_{abc}\epsilon_{ade} (u_b^TC\gamma_5
d_c)(\bar{d}_dC\gamma_5 \bar{s}_e)\ ,
\end{eqnarray}
where the overall negative sign from the identity $\tilde{\Gamma}=-\Gamma$
for $\Gamma=C\gamma_5$ is ignored.

We consider the correlator of the currents to get the QCD sum rule for each meson :
\begin{equation}
\Pi_{S}(q)=i\int d^4x\ e^{iq\cdot x}
\langle0| TJ_{S}(x)J_{S}^{\dagger}(0)|0\rangle\ .
\end{equation}
Within the narrow resonance approximation, including the operators
up to dimension 8, after Borel transforming, we obtain the QCD sum
rules for each meson which can be written in the form,
\begin{eqnarray}
&&C_{0,1}^{S}M^{10}E_4
+C_{4,1}^Sg_c^2\langle G^2\rangle M^6E_2
+\bigg(C_{4,2}^Sm_s\langle\bar{q}q\rangle+C_{4,3}^Sm_s\langle\bar{s}s\rangle\bigg)M^6E_2
\nonumber\\
&&+ \bigg(C_{6,1}^S\langle\bar{q}q\rangle^2
+C_{6,2}^S\langle\bar{q}q\rangle\langle\bar{s}s\rangle\bigg)M^4E_1
\nonumber\\
&&+\bigg(C_{6,3}^Sm_sig_c\langle\bar{s}\sigma\cdot G s\rangle E_1
+m_sig_c\langle\bar{q}\sigma\cdot G q\rangle (C_{6,4}^SE_1+C_{6,5}^S\overline{W}_1)\bigg)M^4
\nonumber\\
&&+m_sg_c^2\langle G^2\rangle
\bigg(\langle\bar{q}q\rangle(C_{8,1}^SE_0+C_{8,2}^SW_0)
+C_{8,3}^S\langle\bar{s}s\rangle E_0\bigg)M^2
\nonumber\\
&&+\bigg(C_{8,4}^S\langle\bar{q}q\rangle
ig_c\langle\bar{s}\sigma\cdot Gs\rangle+C_{8,5}^S\langle\bar{s}s\rangle
ig_c\langle\bar{q}\sigma\cdot Gq\rangle
+C_{8,6}^S\langle\bar{q}q\rangle
ig_c\langle\bar{q}\sigma\cdot Gq\rangle\bigg)M^2E_0
\nonumber\\
&&=2f_{S}^2m_{S}^8e^{-m_{S}^2/M^2}\ ,
\end{eqnarray}
where $M$ is the Borel mass. The decay constant and the mass of
the mesons of the scalar nonet are defined by
\begin{equation}
\langle0|J_S^k|S\rangle=\sqrt{2}f_Sm_S^4\ .
\end{equation}
The contribution from the continuum is encoded in the functions
$E_n(M)$, $W_n(M)$, and $\overline{W}_n(M)$~\cite{Lee05} defined
by
\begin{eqnarray}
E_n(M)&=&\frac{1}{\Gamma(n+1)M^{2n+2}}\int_0^{s_0^2}ds^2\ e^{-s^2/M^2}(s^2)^n\ ,
\nonumber\\
W_n(M)&=&\frac{1}{\Gamma(n+1)M^{2n+2}}\int_0^{s_0^2}ds^2\ e^{-s^2/M^2}(s^2)^n
\bigg(-2\ln(s^2/\Lambda^2)+\ln\pi
\nonumber\\
&&\hspace{4cm}+\psi(n+1)+\psi(n+2)+2\gamma_E-2/3\bigg)\ ,
\nonumber\\
\overline{W}_n(M)&=&\frac{1}{\Gamma(n+1)M^{2n+2}}\int_0^{s_0^2}ds^2\ e^{-s^2/M^2}
(s^2)^n\bigg(-2\ln(s^2/\Lambda^2)+\ln\pi
\nonumber\\
&&\hspace{4cm}+\psi(n+1)+\psi(n+2)+2\gamma_E\bigg)\ ,
\label{sumrule}
\end{eqnarray}
where $s_0$ is the threshold of the continuum and
$\psi(n)=1+1/2+\cdots+1/(n-1)-\gamma_{EM}$.
The first index in the coefficients $C^S_{d,n}$ denotes the
dimension in powers of energy of the operators. In order to
get the contributions from the operators of dimension 6 and 8, we
use the factorization hypothesis which is based on $1/N_c$
arguments. Note that thanks to the structure of the interpolating
currents, the sum rules for $f_0(980)$ and $a_0(980)$ are the
same. The coefficients in the sum rules for each meson are the
following :
\begin{enumerate}

\item{$\sigma$ :}
\begin{equation}
C_{0,1}^{\sigma}=\frac{1}{2^9\cdot 5\pi^6},\
C_{4,1}^{\sigma}=\frac{1}{2^{10}\cdot 3\pi^6},\
C_{6,1}^{\sigma}=\frac{1}{12\pi^2},\
C_{8,6}^{\sigma}=-\frac{1}{12\pi^2},\
\label{Ssigma}
\end{equation}
and the others vanish.

\item{$f_0$ and $a_0$ :}
\begin{eqnarray}
&&C_{0,1}^{f_0,a_0}=\frac{1}{2^9\cdot 5\pi^6},\
C_{4,1}^{f_0,a_0}=\frac{1}{2^{10}\cdot 3\pi^6},\
C_{4,2}^{f_0,a_0}=-\frac{1}{2^4\cdot 3\pi^4},\
C_{4,3}^{f_0,a_0}=\frac{1}{2^5\cdot 3\pi^4},
\nonumber\\
&&C_{6,1}^{f_0,a_0}=0,\
C_{6,2}^{f_0,a_0}=\frac{1}{12\pi^2},\
C_{6,3}^{f_0,a_0}=\frac{1}{2^7\cdot 3\pi^4},\
C_{6,4}^{f_0,a_0}=\frac{1}{2^6\pi^4}=C_{6,5}^{f_0,a_0},\
\nonumber\\
&&C_{8,1}^{f_0,a_0}=-\frac{5}{2^7\cdot 9\pi^4},\
C_{8,2}^{f_0,a_0}=-\frac{1}{2^6\cdot 3\pi^4},\
C_{8,3}^{f_0,a_0}=\frac{1}{2^8\cdot 3\pi^4},\
\nonumber\\
&&C_{8,4}^{f_0,a_0}=-\frac{1}{24\pi^2}=C_{8,5}^{f_0,a_0},\
C_{8,6}^{f_0,a_0}=0\ .
\label{Sf0}
\end{eqnarray}

\item{$\kappa$ :}
\begin{eqnarray}
&&C_{0,1}^{\kappa}=\frac{1}{2^9\cdot 5\pi^6},\
C_{4,1}^{\kappa}=\frac{1}{2^{10}\cdot 3\pi^6},\
C_{4,2}^{\kappa}=-\frac{1}{2^5\cdot 3\pi^4},\
C_{4,3}^{\kappa}=\frac{1}{2^6\cdot 3\pi^4},
\nonumber\\
&&C_{6,1}^{\kappa}=\frac{1}{24\pi^2}=C_{6,2}^{\kappa},\
C_{6,3}^{\kappa}=\frac{1}{2^8\cdot 3\pi^4},\
C_{6,4}^{\kappa}=\frac{1}{2^7\pi^4}=C_{6,5}^{\kappa},\
\nonumber\\
&&C_{8,1}^{\kappa}=-\frac{5}{2^8\cdot 9\pi^4},\
C_{8,2}^{\kappa}=-\frac{1}{2^7\cdot 3\pi^4},\
C_{8,3}^{\kappa}=\frac{1}{2^9\cdot 3\pi^4},\
\nonumber\\
&&C_{8,4}^{\kappa}=-\frac{1}{48\pi^2}=C_{8,5}^{\kappa},\
C_{8,6}^{\kappa}=2C_{8,4}^{\kappa}\ .
\label{Skappa}
\end{eqnarray}

\end{enumerate}

Before finishing this section let us comment on the possible
deviation of the numerical values of the condensates of dimension
6 and 8 from their factorization values. This issue was discussed
in recent papers~\cite{VA,Narison05} using the OPE expansion for
the $V-A$ correlator and data from hadronic $\tau$ decays. It has
been emphasized in~\cite{Narison05} that for the $V-A$ correlator
``due to alternative signs of the condensate contribution in the
OPE and to the fact that in most methods the high-dimension
condensate contributions are corrections to the lowest dimension
condensates in the analysis, the approaches for extracting these
high-dimension condensates can become inaccurate". We point out
that the color and Dirac structure of our condensates of dimension
6 and 8 are different from the analysis of the OPE of the $V-A$
correlator so that it is difficult to use directly their results
in our case. But even if we accept that the ratios of violations
of the factorization hypothesis in our case for the condensates of
dimension 6 and 8 are similar to those presented
in~\cite{Narison05}, our final conclusion will not change due to
the dominant contribution from the condensate of dimension 8 to
the sum rules.

\section{Numerical results}

For the numerical analysis, we use the following values of the
parameters~\cite{Brito05}
\begin{eqnarray}
&&m_s=0.13{\rm GeV},\ \ \langle\bar{u} u\rangle=-(0.23)^3\ {\rm GeV}^3,\ \
f_s=\frac{\langle\bar{s} s\rangle}{\langle\bar{u} u\rangle}=0.8,
\nonumber\\
&&ig_c\langle\bar{q}\sigma\cdot Gq\rangle=0.8\ {\rm
GeV}^2\langle\bar{q} q\rangle, \ \ g_c^2\langle G^2\rangle=0.5\ {\rm
GeV}^4.
\end{eqnarray}
Comparing the strength of the coefficients and the numerical values
of the various condensates, one can see, in the left hand side
(LHS) of the sum rules Eq.~(\ref{sumrule}), that the operators of
dimension 6 and 8 give the main contributions. More precisely the
contributions from the first two operators of dimension 6
operators with the coefficients $C_{6,1}^S$, $C_{6,2}^S$ and the
last three operators of dimension 8 with the coefficients
$C_{8,4}^S$, $C_{8,5}^S$, and $C_{8,6}^S$ dominate the sum rule.
Furthermore, the contribution from the $d=8$ operators comes with
opposite sign to that from the dimension $d=6$ operators in the
physical region of Borel mass $M\approx 1$ GeV, the former
becomes bigger than the latter.

In Figs.~1, 2 and 3, the LHS of the sum rules Eq.~(\ref{sumrule})
as a function of the Borel mass $M$ for $f_0(980)$, $a_0(980)$,
$\sigma(600)$, and $\kappa(800)$ are shown
with the thresholds, $s_0^{f_0}=1.22$ GeV, $s_0^{\sigma}=1.0$ GeV,
and $s_0^{\kappa}=1.1$ GeV~\cite{Brito05,chinese05}, respectively.
\begin{figure}[h]
\begin{minipage}[c]{8cm}
\hspace*{0.5cm} \psfig{file=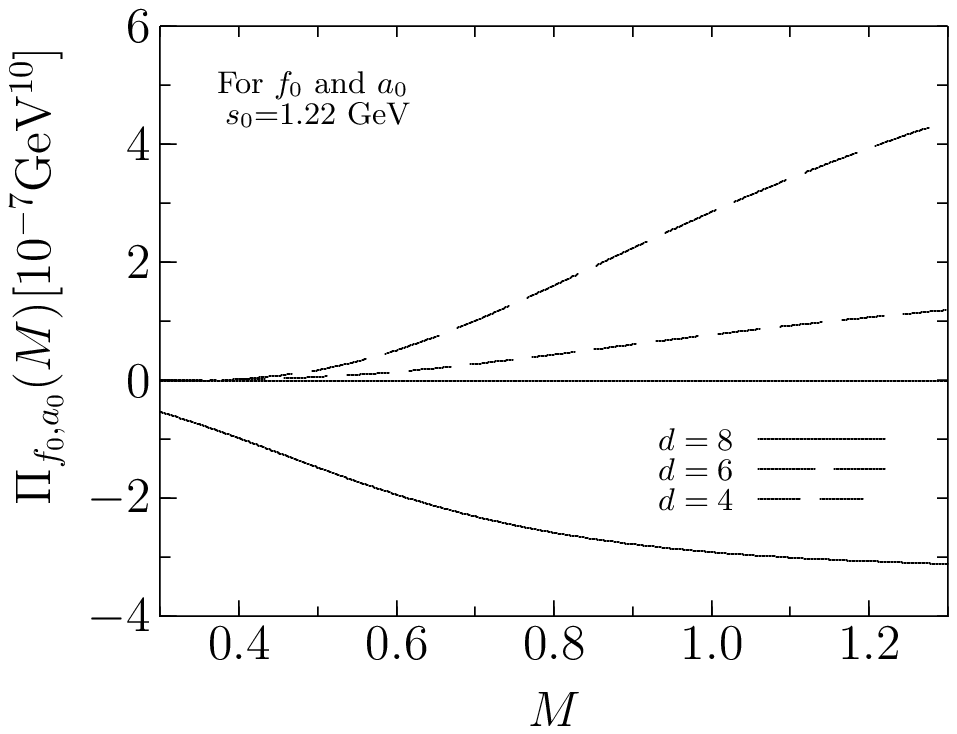,width=6cm,height=5cm}
\caption{The left hand side of the QCD sum rule for $f_0(980)$ and
$a_0(980)$ with the scalar diquark and the scalar antidiquark.}
\end{minipage}
\hspace*{0.5cm}
\begin{minipage}[c]{8cm}
\hspace*{0.5cm} \psfig{file=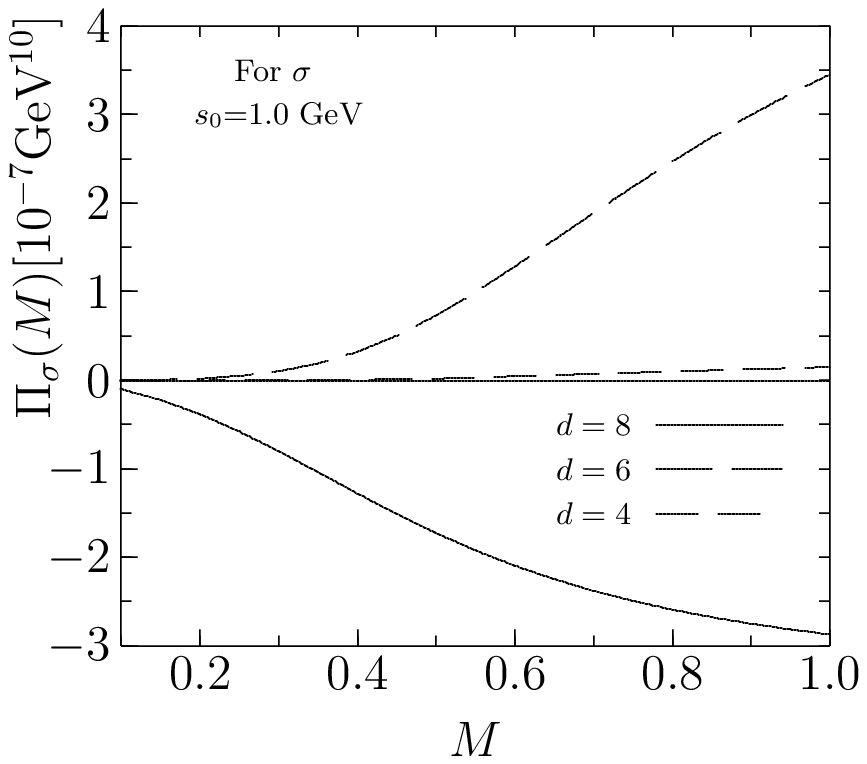,width=6cm,height=5cm}
\caption{The left hand side of the QCD sum rule for $\sigma(600)$
with the scalar diquark and the scalar antidiquark.}
\end{minipage}
\end{figure}

\begin{figure}[h]
\centering
\hspace*{0.5cm} \epsfig{file=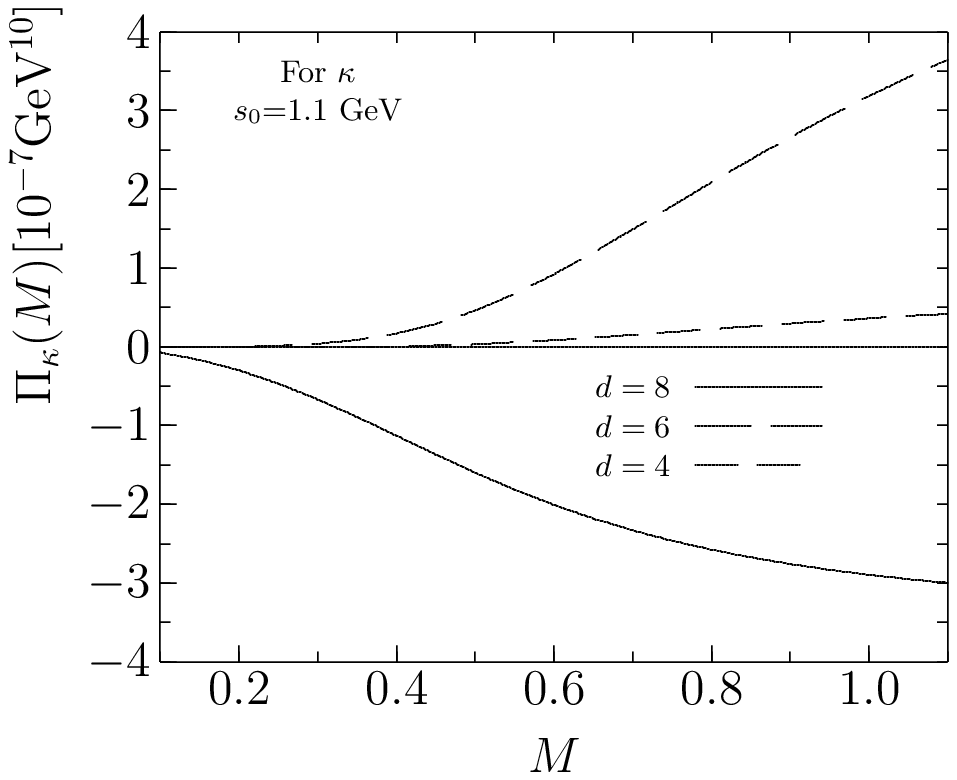,width=6cm,height=5cm}
\caption{The left hand side of the QCD sum rule for $\kappa(800)$
with the scalar diquark and the scalar antidiquark.}
\end{figure}

As shown in Figs.~1, 2, and 3, the most dominant contribution
comes from the operators of dimension 8 : consequently, the QCD
sum rule cannot have physical meaning because the LHS is negative
definite. One could think that contributions from higher
dimensional operators $d>8$ might lead to stabilization of the QCD
sum rules. However, we mention that the contribution from the
operators of dimension 10 to the QCD sum rule is constant. They
have the form of $g_c^2\langle G^2\rangle \langle
\bar{q}q\rangle^2$, $(ig_c\langle\bar{q}\sigma\cdot Gq\rangle)^2$,
and $m_s\langle \bar{q}q\rangle^3$ with the factorization
hypothesis. Since their values are small, their contribution to
the QCD sum rules are expected to be very small. The next
operators are of dimension 12, 14, $\cdots$. They appear with
powers of $M^{-2}$ in the sum rules and therefore their
contribution is expected to be small in region where $M\approx 1$
GeV.

\section{Conclusion}
Our main conclusion is that we do not find a justification for the
interpretation of the light mesons in the scalar nonet as the
scalar diquark--antidiquark bound states within the QCD sum rule
approach. We have demonstrated that the contribution of the
operators of dimension 8 with the factorization hypothesis
is very large and leads to the disappearance of the coupling
of the tetraquark states to the light scalar meson nonet. Of
course, this conclusion might change if another type of
interpolating currents is considered. The investigation of the
properties of tetraquark states within the QCD sum rule approach
with other interpolating diquark currents, e.g., the pseudoscalar
diquark, the vector diquark, or with some mixture of
$q\bar{q}$ configurations, is in progress~\cite{hj}.

\section*{Acknowledgments}
Author is grateful to N.I. Kochelev, V. Vento, H. Kim, and S.H. Lee
for very useful discussions.

\end{document}